\documentclass[aps,prd,onecolumn,superscriptaddress,showpacs,floatfix]{revtex4}
\usepackage{graphicx}
\usepackage{epsfig}
\usepackage{amsmath}
\usepackage{slashed}
\usepackage{amssymb}
\usepackage{units}

\begin{document}

\title{Laser frequency combs and ultracold neutrons to probe braneworlds\\
through induced matter swapping between branes}

\author{Micha\"{e}l Sarrazin}
\email{michael.sarrazin@fundp.ac.be} \affiliation{Department of
Physics,\\ University of Namur (FUNDP),
\\61 rue de Bruxelles, B-5000 Namur, Belgium}

\author{Fabrice Petit}
\email{f.petit@bcrc.be} \affiliation{Belgian Ceramic Research
Centre,\\4 avenue du gouverneur Cornez, B-7000 Mons, Belgium}

\begin{abstract}
This paper investigates a new experimental framework to test the
braneworld hypothesis. Recent theoretical results have shown the
possibility of matter exchange between branes under the influence of
suitable magnetic vector potentials. It is shown that the required
conditions might be achieved with present-day technology. The
experiment uses a source of pulsed and coherent electromagnetic
radiation and relies on the H\"{a}nsch frequency comb technique
well-known in ultrahigh-precision spectroscopy. A good matter
candidate for testing the hypothesis is a polarized ultracold
neutron gas for which the number of swapped neutrons is measured.
\end{abstract}

\pacs{11.25.Wx, 12.60.-i, 13.40.-f, 14.20.Dh}

\maketitle

%
%
%

\section{Introduction}

Over the last two decades, the concept of a braneworld universe has
gained a growing importance in theoretical physics. The braneworld
hypothesis assumes that our Universe is just a membrane (a brane)
embedded in a larger dimensional manifold (a bulk) having $N>4$
dimensions \cite{1,2,3,4}. Standard model particles are expected to
stay confined in our brane while the other branes are normally
considered invisible to us. A wealth of papers have demonstrated
that this geometrical approach offers nice explanations for several
physical phenomena like the dark matter origin \cite{3} or the
hierarchy between the electroweak and Planck scales \cite{4} for
instance. As a consequence, finding evidence of the existence of
branes or extra dimensions is a major challenge for the twenty-first
century. Experimental results could arise from high energy physics
(observation of Kaluza-Klein tower states \cite{5} for instance) or
low energy physics (deviations from the inverse square law of
gravity \cite{6} for instance). In the present paper, we are mainly
motivated by a low energy approach and we explore how the quantum
dynamics of fermions is modified at a nonrelativistic energy scale
when the higher dimensional bulk contains more than only one brane.

In a previous work \cite{7}, it has been shown that for a bulk containing at
least two branes, matter swapping between these two worlds might be possible
(although the effect would remain difficult to observe). In some conditions,
this matter exchange could be triggered by using suitable magnetic vector
potentials \cite{7,8,9}. This effect was studied through a quantum
description of spin-$\nicefrac{1}{2}$ fermions in a $M_4\times Z_2$
universe, which is a model-independent low energy limit of some two-brane
world theories of the Universe \cite{7}. However, the situations considered
in previous papers were rather simplistic \cite{7,8,9}. In the present work,
we investigate further this effect by reconsidering it from a more realistic
experimental point of view. To that end, a simple and inexpensive
experimental setup is proposed, inspired from designs used in neutron
physics investigations \cite{10,11,12,13} and in spectroscopy \cite
{14,15,16,17}.

In section II, the mathematical and physical assumptions underlying
the low energy description of spin-$\nicefrac{1}{2}$ fermions in a
two-brane world are reviewed. The theoretical conditions of matter
swapping between branes are given in section III. We discuss the
environmental conditions that could preclude the swapping to occur.
The role of the magnetic vector potentials, which are required to
match the conditions of successful matter exchange between branes is
also discussed. Finally, in section IV an experimental setup which
might be used to confirm our theoretical prediction is described.
The proposed experiment relies on the use of a polarized ultracold
neutron gas \cite{10,11,12,13} and coherent electromagnetic
radiation thanks to a H\"{a}nsch frequency comblike technique \cite
{14,15,16,17}. It is demonstrated that for certain experimental
parameters, the conditions of a resonant matter exchange between
branes may be obtained.

\section{Model of the low energy limit of two-brane worlds}

\label{Model} In a recent work \cite{7}, regarding the dynamics of spin-$%
\nicefrac{1}{2}$ particles, it has been demonstrated that at low energies
any two-brane world related to a domain wall approach can be described by a
simple noncommutative two-sheeted spacetime $M_4\times Z_2$. For instance, a
two-brane world made of two domain walls on a continuous $M_4\times R_1$
manifold is well modeled by a discrete product space $M_4\times Z_2$ at low
energies. The continuous real extra dimension $R_1$ is replaced by an
effective phenomenological discrete two-point space $Z_2$. At each point
along the discrete extra dimension $Z_2$ there is then a four-dimensional
spacetime $M_4$ endowed with its own metric field. Both branes/sheets are
then separated by a phenomenological distance $\delta $ which is inversely
proportional to the overlap integral of the extra-dimensional fermionic wave
functions of each brane over the fifth dimension $R_1$ \cite{7}. Considering
the electromagnetic gauge field, it has been also demonstrated that the
five-dimensional $U(1)$ bulk gauge field \cite{18} is substituted by an
effective $U(1)\otimes U(1)$ gauge field acting in the $M_4\times Z_2$
spacetime.

It is important to stress that the equivalence between the
continuous two-domain wall approaches and the noncommutative
two-sheeted spacetime model is rather general and does not rely for
instance on the domain walls features or on the bulk dimensionality
\cite{7}. It allows us to conjecture that at low energy, regarding
the quantum dynamics of fermions, any multidimensional setup
containing two branes can be described by a two-sheeted spacetime in
the formalism of noncommutative geometry. Since this matter has
already been considered in detail in a previous work \cite {7}, in
the following, we are only considering the relevant $M_4\times Z_2$
limit.

The mathematical description of fermions in the noncommutative
two-sheeted spacetime is mainly based on the works of Connes
\textit{et al} \cite{19} and relies on the definition of a
noncommutative exterior derivative $D$ acting in $M_4\times Z_2$.
Because of the specific geometrical structure of the bulk, this
operator is given by \cite{7,8,9}:

\begin{equation}
D_\mu =\left(
\begin{array}{cc}
\partial _\mu & 0 \\
0 & \partial _\mu
\end{array}
\right) \text{ and}\ D_5=\left(
\begin{array}{cc}
0 & g \\
-g & 0
\end{array}
\right)  \label{1}
\end{equation}
with $\mu =0,1,2,3$ and where the term $g=1/\delta $ acts as a finite
difference operator along the discrete dimension. $g$ also appears as a
coupling strength, which describes the interaction between the branes.
Although $g$ is a parameter, which results from the exact and complex brane
characteristics, it can be resolute from experiment. One is able to build
the Dirac operator defined as ${\slashed{D}=}\Gamma ^ND_N=\Gamma ^\mu D_\mu
+\Gamma ^5D_5$ by considering the following extension of the gamma matrices
(we are working in the Hilbert space of spinors \cite{19}): $\Gamma ^\mu =%
\mathbf{1}_{\text{2}\times \text{2}}\otimes \gamma ^\mu $\ and\ $\Gamma
^5=\sigma _3\otimes \gamma ^5$, where $\gamma ^\mu $ and $\gamma ^5=i\gamma
^0\gamma ^1\gamma ^2\gamma ^3$ are the usual Dirac matrices and $\sigma _k$ (%
$k=1,2,3$) the Pauli matrices. By introducing a general mass term $M$, a
two-brane Dirac equation is then derived \cite{7}:
\begin{eqnarray}
{\slashed{D}}_{dirac}\Psi &=&\left( {i\slashed{D}-M}\right) \Psi =\left( {%
i\Gamma ^ND_N-M}\right) \Psi =  \label{2} \\
&=&\left(
\begin{array}{cc}
i\gamma ^\mu \partial _\mu -m & ig\gamma ^5-m_c \\
ig\gamma ^5-m_c^{*} & i\gamma ^\mu \partial _\mu -m
\end{array}
\right) \left(
\begin{array}{c}
\psi _{+} \\
\psi _{-}
\end{array}
\right) =0  \nonumber
\end{eqnarray}
where ''$*$'' denotes the complex conjugate. The off-diagonal mass
term $m_c$ can be justified from a two-brane(-domain wall) structure
of the Universe \cite{7}. It can be noticed that by virtue of the
two-sheeted structure of spacetime, the wave function $\psi $ of the
fermion is split into two components, each component living on a
distinct spacetime sheet (i.e. brane). In this notation, the indices
''$+$'' and ''$-$'' allow us to discriminate between the two branes.
If one considers the $(+)$ sheet as our own brane, then the $(-)$
sheet can be considered as a hidden brane.

\subsection{Electromagnetic gauge field}

\label{Gauge}

As explained in the introduction, we want to show how the
electromagnetic field influences the dynamics of fermions. The
reason for the electromagnetic force, in contrast to the electroweak
or strong force, rests on the choice of a simple gauge group and
just serves our experimental purpose. Incorporating
the electromagnetic field $\slashed{A}$ in the model (${\slashed{D}}%
_A\rightarrow {\slashed{D}}+\slashed{A}$) \cite{7,8,9}, the usual $U(1)$
gauge field must be substituted by an extended $U(1)\otimes U(1)$ gauge
field accounting for the two-brane structure \cite{7}. The group
representation is therefore $G=diag(\exp (-iq\Lambda _{+}),\exp (-iq\Lambda
_{-}))$. According to the gauge transformation rule: $\slashed{A}^{\prime }=G%
\slashed{A}G^{\dagger }-iG\left[ {\slashed{D}}_{dirac},G^{\dagger }\right] $%
, the appropriate gauge field is given by (see Ref. \cite{7})
\begin{equation}
\slashed{A}=\left(
\begin{array}{cc}
iq\gamma ^\mu A_\mu ^{+} & \gamma ^5\chi \\
\gamma ^5\overline{\chi } & iq\gamma ^\mu A_\mu ^{-}
\end{array}
\right) \text{ with }\chi =\varphi +\gamma ^5\phi  \label{3}
\end{equation}
where $\varphi $ and $\phi $ are the scalar components of the field
$\chi $ and $\overline{\chi }=\gamma ^0\chi ^{\dagger }\gamma ^0$.
If $\chi $ is different from zero, each charged particle of each
brane becomes sensitive to the electromagnetic fields of both branes
irrespective of their localization in the bulk. This kind of exotic
interaction has been considered previously in literature within the
framework of the mirror matter paradigm \cite{20} and is not covered
by the present paper. Moreover, to be consistent with known physics,
at least at low energies, $\chi $ is necessarily tiny (whereas
$qA_\mu ^{\pm }$ need not to be). This is theoretically corroborated
by the gauge transformation rule which shows that
during each gauge transformation, $\left| \varphi \right| $ (respectively $%
\left| \phi \right| $) varies with an amplitude of order $g$ (respectively $%
\left| m_c\right| $) whatever $\Lambda _{+}$ and $\Lambda _{-}$. Using the
covariant derivative ${\slashed{D}}_A\rightarrow {\slashed{D}}+\slashed{A}$
and according to expression (\ref{3}), the electromagnetic field can be
easily introduced in the two-brane Dirac equation (Eq. (\ref{2})). Then, we
get:
\begin{eqnarray}
\left( {i\slashed{D}}_A{-M}\right) \Psi &=&0  \label{4} \\
&=&\left(
\begin{array}{cc}
i\gamma ^\mu (\partial _\mu +iqA_\mu ^{+})-m & i\widetilde{g}\gamma ^5-%
\widetilde{m}_c \\
i\widetilde{g}^{*}\gamma ^5-\widetilde{m}_c^{*} & i\gamma ^\mu (\partial
_\mu +iqA_\mu ^{-})-m
\end{array}
\right) \left(
\begin{array}{c}
\psi _{+} \\
\psi _{-}
\end{array}
\right) =0  \nonumber
\end{eqnarray}
with $\widetilde{g}=g+\varphi $ and $\widetilde{m}_c=m_c-i\phi $. It is
important to underline that the field $\chi $ just leads to replacing $g$ and $%
m_c$ by the effective parameters $\widetilde{g}$ and $\widetilde{m}_c$. In
the following, we assume that $\widetilde{g}\approx g$ and $\widetilde{m}%
_c\approx m_c$ since $\left| \varphi \right| $ (respectively $\left| \phi
\right| $) should not exceed the amplitude of $g$ (respectively $\left|
m_c\right| $). This choice allows a further simplification of the model. It
is somewhat equivalent to set the off-diagonal term $\chi $ to zero. With
such a choice, we simply assume that the electromagnetic field of a brane
couples only with the particles belonging to the same brane. Each brane
possesses its own current and charge density distribution as sources of the
local electromagnetic fields. On the two branes live then the distinct $%
A_\mu ^{+}$ and $A_\mu ^{-}$ electromagnetic fields. The photon fields $%
A_\mu ^{\pm }$ behave independently of each other and are totally
trapped in their original brane in accordance with observations:
photons belonging to a given brane are not able to reach the other
brane. As a noticeable consequence, the structures belonging to the
branes are mutually invisible by local observers.

\subsection{Nonrelativistic limit and phenomenology}

\label{nonrela}

As explained in the introduction, we are concerned about phenomena
occurring at the nonrelativistic energy scale. In Refs. \cite{8}, it
was indeed demonstrated that any relativistic particle is trapped in
its own brane. As a consequence, the model predicts that there is no
hope to observe the exchange of standard model particles between
branes for relativistic energies $-$ a conclusion that contrasts
with the usual belief on the energy scales at which extra
dimensions' effects should become noticeable. This is the reason why
this paper is restricted to the case of nonrelativistic particles
only.

Let us derive the nonrelativistic limit of the two-brane Dirac equation (%
\ref{4}). Following the well-known standard procedure, a two-brane Pauli
equation can be derived:
\begin{equation}
i\hbar \frac \partial {\partial t}\left(
\begin{array}{c}
\psi _{+} \\
\psi _{-}
\end{array}
\right) =\left\{ \mathbf{H}_0+\mathbf{H}_{cm}+\mathbf{H}_c\right\} \left(
\begin{array}{c}
\psi _{+} \\
\psi _{-}
\end{array}
\right)  \label{5}
\end{equation}
where $\psi _{+}$ and $\psi _{-}$ correspond to the wave functions in the $%
(+)$ and $(-)$ branes respectively. $\psi _{+}$ and $\psi _{-}$ are here
Pauli spinors. The Hamiltonian $\mathbf{H}_0$ is a block-diagonal matrix
where each block is simply the classical Pauli Hamiltonian expressed in each
brane:
\begin{equation}
\mathbf{H}_{\pm }=-\frac{\hbar ^2}{2m}\left( \mathbf{\nabla }-i\frac q\hbar
\mathbf{A}_{\pm }\right) ^2+g_s\mu \frac 12\mathbf{\sigma \cdot B}_{\pm
}+V_{\pm }  \label{6}
\end{equation}
such that $\mathbf{A}_{+}$ and $\mathbf{A}_{-}$ correspond to the
magnetic vector potentials in the branes $(+)$ and $(-)$
respectively. The same convention is applied to the magnetic fields
$\mathbf{B}_{\pm }$ and to the potentials $V_{\pm }$. $g_s\mu $ is
the magnetic moment of the particle with $g_s$ the gyromagnetic
factor and $\mu $ the magneton. In addition to these ``classical''
terms, the two-brane Hamiltonian comprises also new terms specific
to the two-brane world:
\begin{equation}
\mathbf{H}_c=\left(
\begin{array}{cc}
0 & m_cc^2 \\
m_c^{*}c^2 & 0
\end{array}
\right)  \label{7}
\end{equation}
and
\begin{eqnarray}
\mathbf{H}_{cm}=-igg_s\mu \frac 12\left(
\begin{array}{cc}
0 & \mathbf{\sigma \cdot }\left\{ \mathbf{A}_{+}-\mathbf{A}_{-}\right\} \\
-\mathbf{\sigma \cdot }\left\{ \mathbf{A}_{+}-\mathbf{A}_{-}\right\} & 0
\end{array}
\right)  \label{8}
\end{eqnarray}
$\mathbf{H}_c$ is responsible for free spontaneous oscillations of
fermions between each brane. Nevertheless, as mentioned in Refs.
\cite{7,9}, these oscillations are expected to be hardly observable
due to environmental effects which freeze the oscillations (see also
section \ref{freezing}). As a consequence, $\mathbf{H}_c$ will be
neglected in the following \cite{7}. By contrast $\mathbf{H}_{cm}$
is a geometrical coupling involving the gauge fields of the two
branes. It is worth noticing that this coupling term depends on the
magnetic moment and on the difference between the local (i.e. on a
brane) values of the magnetic vector potentials (see Refs.
\cite{7,8,9} for more details and discussions). The coupling term
$\mathbf{H}_{cm}$ implies that matter exchange should be possible
between the branes. Its form suggests that particles could
experience Rabi-like oscillations between the branes that might be
triggered by suitable magnetic vector potentials (see Fig. 1).

\begin{figure}[t]
\centerline{\ \includegraphics[width=4cm,height=4.4cm]{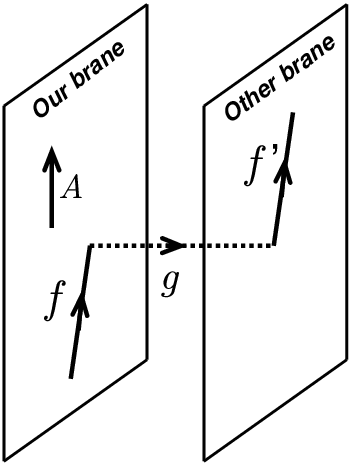}}
\caption{Naive view of the induced transfer of a fermion (a neutron
for instance) between two branes. The fermion \textit{f} in our
brane is transferred to another brane (fermion \textit{f}$^{\prime
}$) under the influence of a suitable magnetic vector potential
$\mathbf{A}$. $g$ represents the coupling constant between branes.}
\label{fig1}
\end{figure}

\section{Resonant matter oscillations between branes}

\label{Swapping}

We consider the possibility of a matter exchange (or swapping) between two
branes through an idealized case and from equations (\ref{5}) to (\ref{8}).
Moreover, $\mathbf{H}_c$ is neglected relative to $\mathbf{H}_{cm}$. Let us
consider a neutral particle ($q=0$) endowed with a magnetic moment (a
neutron for instance), initially ($t=0$) localized in our brane in a region
of curlless rotating magnetic vector potential such that $\mathbf{A}_{+}=%
\mathbf{A}_p=A_p\mathbf{e(}t\mathbf{)}$ and $\mathbf{A}_{-}=0$, with $%
\mathbf{e}(t)=\left( \cos \omega t,\sin \omega t,0\right) $. $\omega $ is
the angular frequency of the field $\mathbf{A}_{+}$, which can be null ($%
\omega =0$) in the static field case. Assuming that the conventional part of
the Pauli Hamiltonian $\mathbf{H}_{\pm }$ can be written as $\mathbf{H}_{\pm
}=V_{\pm }$, any particle initially in a spin-down state for instance
(according to $\mathbf{e}_3=\mathbf{e}_z=\left( 0,0,1\right) $) and
localized in our brane at $t=0$ can be detected in the second brane at time $%
t$ according to the probability \cite{9}:

\begin{equation}
P(t)=\frac{4\Omega _p^2}{(\eta -\omega )^2+4\Omega _p^2}\sin ^2\left( (1/2)%
\sqrt{(\eta -\omega )^2+4\Omega _p^2}t\right)  \label{9}
\end{equation}
where $\Omega _p=gg_s\mu A_p/(2\hbar )$ and $\eta =(V_{+}-V_{-})/\hbar $. In
addition, in the second brane, the particle is then in a spin-up state. $%
\eta \hbar $ is an effective potential that sums up all the interactions
between the particle and its environment described by $\mathbf{H}_{\pm }$.
Those environmental effects are discussed in section \ref{freezing}. Eq. (%
\ref{9}) shows how the particle is transferred to the other brane
through a process involving Rabi-like oscillations. Eq. (\ref{9})
shows that a resonant exchange occurs whenever the magnetic vector
potential rotates with an angular frequency $\omega =\eta $. A
similar expression is obtained by applying the substitution $\omega
\rightarrow -\omega $ if the particle is in a spin-up state at $t=0$
(the resonance is then achieved with a counterrotative vector
potential). In this case, the particle is in a spin-down state in
the second brane after exchange. Note that Eq. (\ref{9}) corresponds
to a resonant process with a half-height width $\Delta \omega
=4\Omega _p$. The weaker the coupling constant $g$ is, the narrower
the resonance is. By contrast, the greater $A_p$ is, the broader the
resonance is.

Of course when $\omega \neq \eta $, or even in the static field case ($%
\omega =0$), matter oscillations between branes occurs as well.
Nevertheless, due to environmental effects discussed in section \ref
{freezing}, the amplitude of these oscillations must probably be strongly
damped and hardly observable.

It is clear that the situation described by Eq. (\ref{9}) remains rather
simplistic. In the suggested experiment (section \ref{Experiment}) we will
propose a more realistic situation to achieve resonant matter exchange
through the use of coherent electromagnetic radiations.

\subsection{Freezing oscillations and environmental effective potential $%
\eta $}

\label{freezing}

As detailed in previous papers \cite{8,9}, usual environmental interactions
are strong enough to suppress matter oscillations between adjacent branes.
This can be easily checked by considering Eq. (\ref{9}) showing that if $%
\omega \neq \eta $ then $P(t)$ decreases as $\eta $ increases in
comparison to $\Omega _p$. Based on the fact that no such
oscillations have been observed so far, we can expect that the ratio
$\Omega _p/\eta $ is usually very small. Therefore, any experiment
seeking for such oscillations should focus on the quest of resonant
responses (i.e. when $\omega =\eta $) in order to avoid the
environmental confinement.

Next, if one considers an experiment involving a set of particles
with strong collisional dynamics, any coherent oscillatory behavior
would probably be inhibited. From the point of view of a single
particle, each collision resets the probability of transfer
according to a quantum Zeno-like effect. This oscillation damping is
also expected to increase dramatically with temperature.

Note that the same environmental effects are responsible for the negligible
role of $\mathbf{H}_c$, which cannot be artificially enhanced by contrast to
$\mathbf{H}_{cm}$.

Therefore a prerequisite for observing the oscillatory behavior of the
particles between branes is to keep each particle isolated from the
environment as much as possible and to apply very specific magnetic vector
potentials, i.e. a context never met in any kind of experiments so far.

As a consequence, it seems natural to work preferentially with
ultracold neutrons \cite{10,11,12,13}, which are insensitive to
electric fields. In addition, neutron magnetic sensibility is the
result of its magnetic moment only. Therefore, convenient Helmholtz
coils should be used to cancel the ambient magnetic fields (such as
the Earth's). Ambient electromagnetic waves could be cancelled too
in a large spectral frequency domain by working with a cooled setup
(to avoid black body emissions) and using a combination of
convenient shields, such as the Faraday cage and lead brick walls.
Moreover, one may assume that by using a low-density neutron gas the
collisions between particles should be prevented. Of course, such an
experimental setup should be emptied of atmospheric gases.

At last, note that for a neutron shielded from magnetic fields, $\eta
=(V_{grav,+}-V_{grav,-})/\hbar $ (i.e. only gravitational contributions are
considered). However, it is difficult to assess the value of $\eta \hbar $.
For instance, the estimations given in Ref. \cite{9} suggest that $%
V_{grav,+} $ could be of the order of $-500$ eV owing to the Milky Way core
gravitational influence exerted on neutrons. By contrast, the Sun, the Earth
and the Moon lead respectively to contributions of about $-9$ eV, $-0.65$ eV
and $-0.1$ meV. Nevertheless, since the gravitational contribution of the
other brane ($V_{grav,-}$) remains unknown, $\eta $ appears therefore as an
effective unknown parameter (which can be either positive or negative) of
the model. In addition, $\eta $ might be also time-dependent. It is
instructive for instance to consider the motion of Earth around the Sun.
From the aphelion to the perihelion, the gravitational energy (due to the
Sun) of a neutron varies from $-9.12$ eV to $-9.43$ eV leading then to an
absolute shift of $\eta $ of about $1.7$ meV/day. The time dependence of $%
\eta $ could have several different origins such as the relative
particle motion with reference to the unknown mass distribution in
the second brane for instance. It is therefore impossible to
theoretically assess the time rate of shift of $\eta $ even if this
value is expected to be very small. As a consequence, the unknown
magnitude and time dependence of $\eta $ are among the main
difficulties in the search for resonant matter swapping between
branes.

\subsection{Magnetic vector potential: Ambient and artificial contributions}

\label{Magvect}

Before detailing a suitable experimental setup, we first need to
address the issue of a hypothetical ambient magnetic vector
potential. The existence of such a potential was recently debated in
literature in the context of photon mass measurement \cite{21}. For
instance, in Refs. \cite{21}, ambient
magnetic vector potential was estimated by integrating $\mathbf{B}_{amb}%
\mathbf{=\nabla \times A}_{amb}$ and considering ambient magnetic fields $%
\mathbf{B}_{amb}$. It was demonstrating that $A_{amb}=RB_{amb}$,
where $R$ is the typical distance from sources. From earth magnetic
field measurements, it was therefore assessed there is an ambient
contribution $A_{amb}$ of about $200$ T$\cdot $m whereas from the
Coma galactic cluster magnetic field, a value of $A_{amb}\sim
10^{12}$ T$\cdot $m was calculated. Although these assessments do
not really constrain the present model, they raise a number of
questions. Indeed, several authors have pointed out that any
assessment of the ambient magnetic vector potential from such simple
calculations was unreliable (see Luo \textit{et al} \cite{21}). The
magnetic vector potential derived that way is the transversal
contribution only whereas any vector potential comprises three parts: $\mathbf{A}_{amb}\mathbf{%
=A}_{\bot }\mathbf{+A}_{\Vert }\mathbf{+A}_{cte}$. Here $\mathbf{A}_{\bot }$
is the transverse part such that $\mathbf{B}_{amb}\mathbf{=\nabla \times A}%
_{\bot }$, $\mathbf{A}_{\Vert }$ is the longitudinal component such that $%
\mathbf{\nabla \times A}_{\Vert }=\mathbf{0}$, and $\mathbf{A}_{cte}$ is a
constant vector. It is obvious that a rigorous assessment of the ambient
magnetic vector potential has to take into account all these components.
Unfortunately they cannot be determined since several boundary conditions
are still unknown due to their astrophysical origins. In addition, Eq. (\ref
{8}) shows that it is the net difference $\delta \mathbf{A}=\mathbf{A}_{+}-%
\mathbf{A}_{-}$ between the vector potentials of the two branes that
is relevant, not the local values on the branes. For all these
reasons, the effective ambient magnetic vector potential $\delta
\mathbf{A}_{amb}$
''could be very large... or null.'' To simplify, we consider hereafter $%
\delta \mathbf{A}_{amb}$ as an unknown parameter of the model. Nevertheless,
to be consistent with observations, Eq. (\ref{9}) shows that any ambient
magnetic vector potential $\delta \mathbf{A}_{amb}$ should be balanced by an
ambient confining potential $\eta \hbar $, such that $\Omega _p/\eta \ll 1$
(in order to avoid unseen oscillations of ''free'' fermions).

Finally, since in the following one is seeking for a resonant oscillatory
mechanism, a first requirement is to consider a rotating magnetic vector
potential as explained before. Such a potential can obviously be obtained
from a coherent electromagnetic wave with a circular polarization. In that
case, the magnetic vector potential $\mathbf{A}$ is trivially related to the
electric and magnetic fields through $\mathbf{E}=-\partial \mathbf{A}%
/\partial t$ and $\mathbf{B=\nabla \times A}$.

\section{Setup for resonant neutron swapping between branes}

\label{Experiment}

In the present section, we give a sketch of an experimental setup that might
be suitable to investigate the resonant mechanism described in the previous
section. Under a suitable rotative field $\mathbf{A}_p$, a neutron $n$ may
disappear from our brane to the other one, i.e. $n\rightarrow n^{\prime }$
(where the ''prime'' denotes the neutron in the second brane), but for an
observer in our brane we get $n\rightarrow $ $nothing$ (Fig. 1). In the
present experiment, we consider an ultracold neutron gas (i.e. neutrons with
a kinetic energy around $100$ neV) as justified in section \ref{freezing}.

In a Coulomb gauge, a rotative magnetic vector potential $\mathbf{A}_p$ can
be obtained from a circularly polarized electromagnetic wave. However, to
achieve the conditions of a successful matter exchange, one must deal with
the following difficulties:

(i) The value of the coupling constant $g$ likely results in hardly
detectable effects. If the coupling constant $g$ is small enough,
the phenomenon will not be easily observed since the resonance width
will then become narrower than the bandwidth of any available
coherent sources. To achieve the conditions of a successful matter
exchange, the applied electromagnetic waves must possess a bandwidth
lower than the resonance width, as shown in section \ref{Swapping}.
In Refs. \cite{7,8,9}, values such that $g\leq 10^3$ m$^{-1}$ were
usually considered, though it was shown that our model could be in
agreement with known physics even for $g<10^{10}$ m$^{-1}$\cite{8}.
However those values are very low and remain problematic in the
present experimental context.

(ii) The unknown magnitude and time dependence of $\eta $ make
difficult or even preclude the observation of a resonant swapping at
a fixed frequency.

To overcome these restrictions, a possibility is to consider the
frequency comb technique introduced by H\"{a}nsch \cite{14} and well
known by spectroscopists \cite{15,16,17}: excitation by a train of
phase-coherent pulses allows us to reach narrow resonance bandwidths
with a larger efficiency on a large wavelength domain. Such a
frequency comb source can be produced by using mode-locked
(phase-locked) lasers through the multiple reflections of a pulse
inside an optical Fabry-P\'{e}rot cavity. Nevertheless, we do not
describe or discuss here the detailed way to produce a frequency
comb source, which has been widely described elsewhere by many other
authors \cite {14,15,16,17} and which is out of the scope of the
present paper.

\begin{figure}[t]
\centerline{\ \includegraphics[height=5cm]{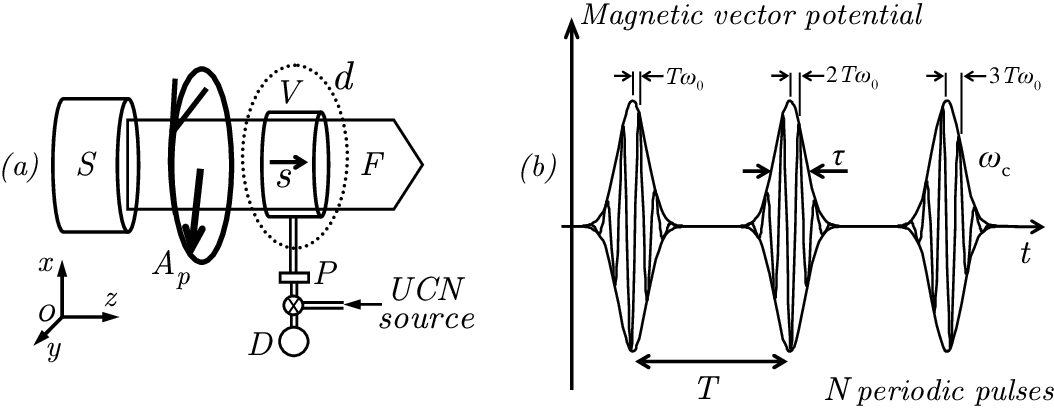}}
\caption{Sketch of the experimental device for the study of particle
swapping between branes. a: In a first step, the ultracold neutron
gas is stored in the vessel ($V$) after neutrons have been polarized
by the polarizer ($P$). Neutrons' spins ($s$) are polarized
perpendicularly to the plane of rotation of $\mathbf{A}_p$. In a
second step, the neutron gas is subjected to a pulsed
electromagnetic beam ($F$) from a frequency comb source ($S$). After
a run of excitations, the vessel is emptied and neutrons are counted
with a detector ($D$). The device is completed by a neutron decay
detector ($d$). b: Sketch of the amplitude of the magnetic vector
potential related to the beam ($F$) against time $t$. The field is a set of $%
N$ coherent pulses. $T$ is the time period between each pulse while $\tau $
is the pulse duration. $\omega _c$ is the pulsation of the carrier wave
while $T\omega _0$ denotes the dephasing between the carrier wave and the
pulse maximum.}
\label{fig2}
\end{figure}

Let us now study the applicability of this technique to force the swapping
of neutrons between the branes.

\subsection{Source of excitation}

Ultracold neutrons are stored in a convenient vessel (emptied of
atmospheric gases) which is supplied with a train of $N$
phase-coherent
pulses related to an electromagnetic wave having a magnetic field amplitude $%
B_0$ (see Fig. 2). The magnetic vector potential felt by a neutron in the
cavity can be expressed as \cite{14}:
\begin{equation}
\mathbf{A}_p(t)=\frac{B_0c}{\omega _c}\sum_{n=0}^{N-1}\Upsilon _\tau (t-nT)%
\mathbf{e(}t-nT_0\mathbf{)}  \label{10}
\end{equation}
where $\mathbf{e(}t\mathbf{)=}\left( \cos \omega _ct,\sin \omega
_ct,0\right) $, and $\omega _c$ is the angular frequency of the
carrier wave. $\Upsilon _\tau (t)$ is the envelope function of a
pulse, with $\tau $ the temporal pulse-width (see Fig. 2.b). One
considers a usual Gaussian pulse such that $\Upsilon _\tau (t)=$
$\exp (-4(t/\tau )^2\ln 2)$. $T$ is the time period between each
pulse (see Fig. 2.b) such that $T\gg \tau $. One defines the
frequency of repetition $f_r=1/T$ , with $2\pi f_r\ll \omega _c$.
$T_0$ allows a phase difference between the carrier wave and the
pulse maximum, with $\left| T-T_0\right| \leq 2\pi /\omega _c$. The
spectrum of the field is therefore given by
\begin{equation}
\left| \mathbf{A}_p(\omega )\right| ^2=\frac{B_0^2c^2}{2\omega _c^2}\left\{
S(\omega +\omega _0)\overline{\Upsilon }_\tau ^2(\omega +\omega _c)+S(\omega
-\omega _0)\overline{\Upsilon }_\tau ^2(\omega -\omega _c)\right\}
\label{10bis}
\end{equation}
with
\begin{equation}
\overline{\Upsilon }_\tau (\omega )=\int \Upsilon _\tau (t)e^{i\omega t}dt=%
\frac{\tau \sqrt{\pi }}{2\sqrt{\ln 2}}\exp \left( -\frac{\omega ^2\tau ^2}{%
16\ln 2}\right)  \label{10ter}
\end{equation}
and
\begin{equation}
S(\omega )=\left| \sum_{n=0}^{N-1}e^{i\omega Tn}\right| ^2=\frac{\sin
^2(\omega TN/2)}{\sin ^2(\omega T/2)}  \label{10qua}
\end{equation}
and where
\begin{equation}
\omega _0=\omega _c\left( 1-\frac{T_0}T\right)  \label{10pen}
\end{equation}
Such a spectrum is a frequency comb (see Fig. 3) made of narrow
peaks located at $\omega _k=2k\pi f_r+\omega _0$ ($k\in\mathbb{N}$)
for $\omega >0$ (or at $\omega _k=-2k\pi f_r-\omega _0$
($k\in\mathbb{N}$) for $\omega <0$). For these frequencies one gets
$S(\omega _k)=N^2$. The available frequencies can be fine-tuned by
adjusting $\omega _0$ (or $f_r$). The width at half-height of each
peak is $\delta \omega \sim 5.57f_r/N$, such that the temporal
coherence of each frequency rises as $N$ increases. One defines the
fineness $F=2\pi f_r/\delta \omega \sim N$. The number $Q$ of useful
frequency peaks is given by the width at half-height $\Delta \omega
$ of the squared Gaussian function, i.e. $\Delta \omega
=4\sqrt{2}\ln 2/\tau \sim 4/\tau $ and one gets $Q=\Delta \omega
/(2\pi f_r)=(2/\pi )(T/\tau )$.

\begin{figure}[t]
\centerline{\ \includegraphics[width=8cm]{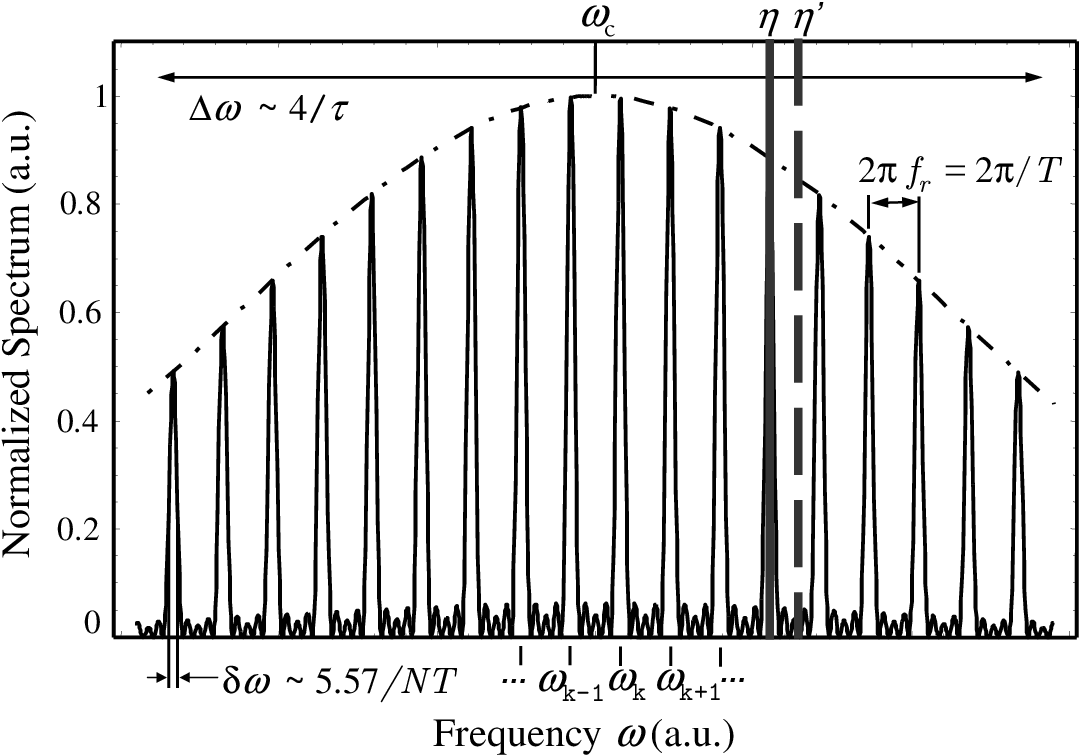}} \caption{Naive
view of the frequency comb related to the spectrum of the pulsed
coherent source. The full band ($\eta $) and the dashed band ($\eta
^{\prime }$) correspond to two different possible values of the
effective potential. The thickness of the bands is about $\delta
\omega $ and
corresponds to the width of the resonances. $\eta ^{\prime }$ can shift towards $%
\eta$ provided a suitable time variation of the potential. If the
potential is not time-dependent, $\eta ^{\prime }$ can match the
frequency of a tooth of the comb provided that $\omega _0$ is
fine-tuned. The real number of teeth is many thousand times greater
than on the present drawing.} \label{fig3}
\end{figure}

Since $\mathbf{E}_p=-\partial \mathbf{A}_p/\partial t$, it must be noticed
that the electric field of the wave presents a different pulse shape from
the magnetic vector potential: the wave spectra considered from the electric
and vector potential fields point of view are related through $\left|
\mathbf{E}_p(\omega )\right| ^2=\omega ^2$ $\left| \mathbf{A}_p(\omega
)\right| ^2$. This point is important since the pulse shape control can be
experimentally obtained by properly designing the electric field \cite{22}.

In the experiment, any neutron is also submitted to the magnetic field $%
\mathbf{B}_p=(1/c)\mathbf{e}_z\times \mathbf{E}_p$ of the electromagnetic
wave such that we get:

\begin{equation}
\mathbf{B}_p\mathbf{(}t\mathbf{)}=B_0\sum_{n=0}^{N-1}\Upsilon _\tau (t-nT)%
\mathbf{e}_m^{\left( n\right) }\mathbf{(}t\mathbf{)}  \label{11ante}
\end{equation}
where $\mathbf{e}_m^{(n)}(t)=\mathbf{e}(t-nT_0)\mathbf{+}u(t-nT)\ \mathbf{e}%
_z\times \mathbf{e}(t-nT_0)$ with $u(t)=8t\ln 2/(\omega _c\tau ^2)$ and $%
\mathbf{e}_z=(0,0,1)$.

\subsection{Neutron dynamics}

In the experiment, in addition to the magnetic vector potential $\mathbf{A}%
_p(t)$, neutrons are submitted to the ambient gravitational fields ($V_{\pm
} $) and the magnetic field $\mathbf{B}_p(t)$ of the electromagnetic wave.
For completeness, we should also consider an ambient magnetic vector
potential $\delta \mathbf{A}_{amb}=\delta A_{amb}\mathbf{u}$, where $\mathbf{%
u}=\left( \sin \theta \cos \varphi ,\sin \theta \sin \varphi ,\cos \theta
\right) $ in the system of reference $Oxyz$ (see Fig. 2.a). The low energy
of neutrons in the vessel makes the nonrelativistic equations (\ref{5}) to (%
\ref{8}) valid. Therefore, from Eqs. (\ref{10}) and (\ref{11ante}), Eq. (\ref
{5}) can be expressed as
\begin{eqnarray}
i\hbar \frac \partial {\partial t}\left| \Psi \right\rangle &=&\left\{
\left[
\begin{array}{cc}
V_{+} & -iE_a\mathbf{\sigma \cdot u} \\
iE_a\mathbf{\sigma \cdot u} & V_{-}
\end{array}
\right] +E_0\sum_{n=0}^{N-1}\Upsilon _\tau (t-nT)\times \right.  \nonumber \\
&&\left. \left[
\begin{array}{cc}
\mathbf{\sigma \cdot e}_m^{(n)}\mathbf{(}t\mathbf{)} & -i\kappa \mathbf{%
\sigma \cdot e(}t-nT_0\mathbf{)} \\
i\kappa \mathbf{\sigma \cdot e(}t-nT_0\mathbf{)} & 0
\end{array}
\right] \right\} \left| \Psi \right\rangle  \label{11}
\end{eqnarray}
with $E_0=(1/2)g_s\mu B_0$, $\kappa =gc/\omega _c$ and $E_a=(1/2)gg_s\mu
\delta A_{amb}$.

According to Eq. (\ref{11}), if $E_0=E_a=0$, in our brane, a neutron can be
described by the eigenstates $\left| \psi _{\pm 1/2}^{+}\right\rangle $ with
energy $V_{+}$ (the subscript ''$\pm 1/2$'' denotes both spin states along
the $Oz$ axis), whereas in the second brane, it will be described by the
eigenstates $\left| \psi _{\pm 1/2}^{-}\right\rangle $ with energy $V_{-}$.
In Eq. (\ref{11}), it is then possible to write
\begin{eqnarray}
\left| \Psi \right\rangle &=&a(t)e^{-i\hbar ^{-1}V_{+}t}\left| \psi
_{+1/2}^{+}\right\rangle +b(t)e^{-i\hbar ^{-1}V_{+}t}\left| \psi
_{-1/2}^{+}\right\rangle  \label{12} \\
&&+c(t)e^{-i\hbar ^{-1}V_{-}t}\left| \psi _{+1/2}^{-}\right\rangle
+d(t)e^{-i\hbar ^{-1}V_{-}t}\left| \psi _{-1/2}^{-}\right\rangle  \nonumber
\end{eqnarray}
Substituting (\ref{12}) into (\ref{11}), the Pauli equation (\ref{11})
reads:
\begin{equation}
\left\{
\begin{array}{c}
i\hbar \partial _ta(t)= \\
i\hbar \partial _tb(t)= \\
i\hbar \partial _tc(t)= \\
i\hbar \partial _td(t)=
\end{array}
\begin{array}{c}
E_0b(t)e^{-i\omega _ct}(\Lambda (t)-i\widetilde{\Lambda }(t))-i\kappa
E_0d(t)e^{i(\eta -\omega _c)t}\Lambda (t)-iE_ac(t)\cos \theta e^{i\eta
t}-iE_ad(t)e^{-i\varphi }\sin \theta e^{i\eta t} \\
E_0a(t)e^{i\omega _ct}(\Lambda ^{*}(t)+i\widetilde{\Lambda }^{*}(t))-i\kappa
E_0c(t)e^{i(\eta +\omega _c)t}\Lambda ^{*}(t)+iE_ad(t)\cos \theta e^{i\eta
t}-iE_ac(t)e^{i\varphi }\sin \theta e^{i\eta t} \\
i\kappa E_0b(t)e^{-i(\eta +\omega _c)t}\Lambda (t)+iE_aa(t)\cos \theta
e^{-i\eta t}+iE_ab(t)e^{-i\varphi }\sin \theta e^{-i\eta t} \\
i\kappa E_0a(t)e^{-i(\eta -\omega _c)t}\Lambda ^{*}(t)-iE_ab(t)\cos \theta
e^{-i\eta t}+iE_aa(t)e^{i\varphi }\sin \theta e^{-i\eta t}
\end{array}
\right.  \label{13}
\end{equation}
with $\Lambda (t)=\sum_{n=0}^{N-1}\Upsilon _\tau (t-nT)e^{i\omega _cnT_0}$
and $\widetilde{\Lambda }(t)=\sum_{n=0}^{N-1}\Upsilon _\tau
(t-nT)u(t-nT)e^{i\omega _cnT_0}$, and where ''$*$'' denotes the complex
conjugate.

Eqs. (\ref{13}) are complex to solve exactly without numerical
computation. Nevertheless, the secular approximation and the first
order perturbation theory can be applied to obtain an approximate
solution of the system (\ref {13}). In the following, we assume that
$\omega _c$ has the same magnitude as $\left| \eta \right| $.

First, if $\kappa =0$ and $E_a=0$, from the first order perturbation theory
it can be shown that the terms involving the function $\Lambda (t)-i%
\widetilde{\Lambda }(t)$ lead to an amplitude contribution equal to zero and
can therefore be ignored in the whole system \cite{23}.

Next, it is worth remembering that Eq. (\ref{9}) also implies
$E_a/\hbar \left| \eta \right| \ll 1$ for consistency (to avoid
undetected ''free'' oscillations of fermions, see section
\ref{Magvect}). Then, considering the secular approximation, the
terms with $E_a\exp (\pm i\eta t)$ exhibiting rapid variations can
be neglected. Indeed these terms lead to amplitude contributions
$E_a/(\hbar \eta )$, which are then much smaller than $1$. It must
be pointed out that the influence of a hypothetic ambient magnetic
vector potential vanishes in that case.

At last, if $\eta >0$ (or $\eta <0$), the terms with $\kappa E_0\exp
(\pm i(\eta +\omega _c)t)$ (or $\kappa E_0\exp (\pm i(\eta -\omega
_c)t)$) exhibit rapid variations since $\left| \eta +\omega
_c\right| \gg \left| \eta -\omega _c\right| $ (or $\left| \eta
-\omega _c\right| \gg \left| \eta +\omega _c\right| $). As a
consequence, the secular approximation allows us to neglect the
terms with $\exp (\pm i(\eta +\omega _c)t)$ if $\eta >0$ (or the
terms with $\exp (\pm i(\eta -\omega _c)t)$ if $\eta <0$).

These simplifications allow us to rewrite the system (\ref{13}) as
two independent systems in a very compact form:

(i) If $\eta >0$:
\begin{equation}
\left\{
\begin{array}{c}
\partial _ta(t)=-\Omega _pe^{i(\eta -\omega _c)t}\Lambda (t)d(t) \\
\partial _td(t)=\Omega _pe^{-i(\eta -\omega _c)t}\Lambda ^{*}(t)a(t)
\end{array}
\right.  \label{14}
\end{equation}

(ii) If $\eta <0$:
\begin{equation}
\left\{
\begin{array}{c}
\partial _tb(t)=-\Omega _pe^{i(\eta +\omega _c)t}\Lambda ^{*}(t)c(t) \\
\partial _tc(t)=\Omega _pe^{-i(\eta +\omega _c)t}\Lambda (t)b(t)
\end{array}
\right.  \label{14bis}
\end{equation}
with $\Omega _p=\kappa E_0/\hbar $. For a fixed circular polarization of the
electromagnetic wave, the two systems (\ref{14}) and (\ref{14bis}) introduce
a constraint on the spin state in which the neutron must be prepared to
allow the swapping. Since the influence of the electromagnetic wave can be
treated as a simple perturbation, the systems (\ref{14}) and (\ref{14bis})
can be solved analytically by using first order perturbation theory. For a
neutron initially localized in our brane at time $t=0$, and prepared with an
initial spin state $s=\pm 1/2$ (in relation to $\eta =\pm \left| \eta
\right| $), the probability $P_{\pm 1/2}$ to find the particle in the second
brane is finally:
\begin{equation}
P_{\pm 1/2}=\Omega _p^2\overline{\Upsilon }_\tau ^2(\omega _c\mp \eta
)S(\eta \mp \omega _0)  \label{15}
\end{equation}
with $S(\eta \mp \omega _0)=S(\omega =\eta \mp \omega _0)$ (see Eq. (\ref
{10qua})). According to Eqs. (\ref{10bis}), (\ref{10ter}) and (\ref{10qua}),
Eqs. (\ref{14}) to (\ref{15}) show that a resonant swapping occurs when the
frequency $\omega _k=2k\pi f_r\pm \omega _0$ of the frequency comb matches $%
\eta $. In addition, the width at half-height of each resonance is $\delta
\omega \sim 5.57f_r/N$, i.e. corresponds to the width of a tooth of the comb
(Fig. 3). Assuming that the frequency of resonance is achieved in the domain
of frequencies covered by the comb, one gets:
\begin{equation}
P_{\pm 1/2}\sim \tau ^2\Omega _p^2N^2  \label{16}
\end{equation}
As shown by H\"{a}nsch for spectroscopy \cite{14}, the probability
$P_{\pm 1/2}$ increases as the square of $N$ increases; i.e. the
probability increases as the teeth of the frequency comb become
narrower. In addition, the use of a frequency comb technique allows
us then to achieve intense pulsed fields enhancing $\Omega _p$ and
thus $P_{\pm 1/2}$. Indeed, the pulsed mode permits intense
electromagnetic fields and therefore large values of $A_p$. For
instance, common pulsed sources can easily reach intensities of
$10^{14}$ W$\cdot$cm$^{-2}$ up to $10^{18}$ W$\cdot$cm$^{-2}$
\cite{24}. Since $\Omega _p=gg_s\mu A_p/(2\hbar )$, one can increase
the coupling and $P_{\pm 1/2}$ despite the weak value of $g$. The
frequency comb seems therefore to be a better candidate than any
other electromagnetic field configuration for achieving conditions
of a successful matter swapping between branes.

Now, since the excitation time $t_e$ during which neutrons feel a set of
coherent pulses is $t_e\sim NT$, the rate $\Gamma $ of neutron exchange
between branes is simply (at resonance):
\begin{equation}
\Gamma =P_{\pm 1/2}/t_e=\tau ^2\Omega _p^2N/T  \label{17}
\end{equation}

\subsection{Experimental consequences and discussion}

\begin{figure}[t]
\centerline{\ \includegraphics[width=8cm]{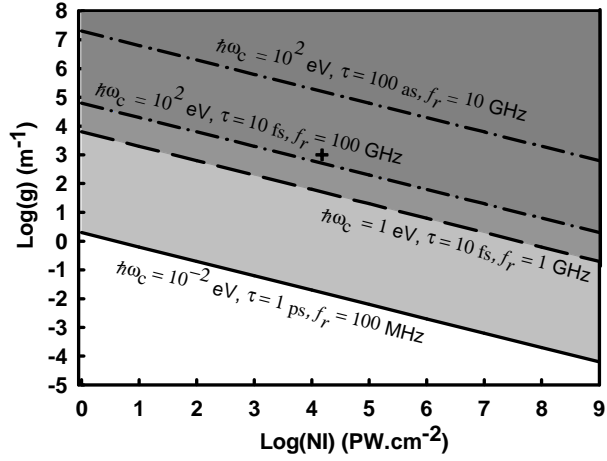}}
\caption{Values of the coupling constant $g$ leading to an observable
swapping of neutrons between two branes. The values of $g$ are given against
the effective pulsed beam intensity $NI$. Each line is the lower reachable
limit, defined by $\Gamma =0.1\Gamma _n$, for some specific carrier wave
pulsation $\omega _c$, pulse duration $\tau $, and frequency repetition $f_r$%
. Grey areas up to a given line correspond to $\Gamma \geq 0.1\Gamma _n$.}
\label{fig4}
\end{figure}

(1) \textbf{Neutron gas polarization}.

According to Eqs. (\ref{14}) and (\ref{14bis}), a neutron initially in a
spin-up state in our brane reaches the other brane in a spin-down state (and
reciprocally). Therefore, in order to achieve a successful transfer, the
neutron gas in the vessel must be polarized before the experiment with a
direction of polarization normal to the plane of rotation of the magnetic
vector potential of the incident electromagnetic wave (Fig. 2.a). It appears
that the collisional dynamics between neutrons or in relation with the
storage chamber walls is negligible provided that collisional time $t_c$ is
greater than $t_e$ which is assumed to be satisfied in the experimental
conditions. Moreover since the neutron gas is polarized, the Pauli exclusion
principle will favor an increase of the collisional time $t_c$ thus
promoting the experimental success.\\

(2) \textbf{Measure of the rate of neutron exchange}.

Let $N_0$ be the initial number of neutrons in the storage chamber.
$\Gamma _n$ is the loss rate of neutrons that takes into account the
usual decay rate and the losses in the vessel (in the following, one
only uses the usual decay rate value for computations). $t_s$ is the
storage time (i.e. the duration of an experiment). In a first
calibration experiment, $N_c(t_s)=N_0\exp (-\Gamma _nt_s)$ will be
the number of remaining neutrons in the vessel after an experimental
run without an applied electromagnetic field. In a second experiment
where the wave is switched on, the number of recorded neutrons will
be then given by: $N_{c,f}(t_s)=N_0\exp (-\Gamma t_s)\exp (-\Gamma
_nt_s)$. Then, the neutron transfer between branes could be simply
detected by measuring the ratio: $N_{c,f}(t_s)/N_c(t_s)=\exp
(-\Gamma t_s)$. The number of stored neutrons, before and after
electromagnetic field switching, can be directly counted (see Fig.
2.a) by a helium-3 neutron detector for instance (see Baker
\textit{et al} \cite{10}).

According to the duration of the experiment, it can be relevant to also
measure the neutron decays. In a first calibration experiment, $%
N_c^{(d)}(t)=N_0(1-\exp (-\Gamma _nt))$ will be then the number of usual
neutron decays during an experimental run without the electromagnetic field.
In a second experiment where the wave is switched on, the number of recorded
events will be then given by: $N_{c,f}^{(d)}(t)=N_0(\Gamma _n/(\Gamma
_n+\Gamma ))\cdot (1-\exp (-(\Gamma _n+\Gamma )t))$. Then, the neutron
transfer between branes could be simply described through the ratio: $%
(dN_{c,f}^{(d)}(t)/dt)/(dN_c^{(d)}(t)/dt)=\exp (-\Gamma t)$.

According to some previous experimental results about ultracold neutrons
\cite{10,11,12,13}, a relevant criterion to clearly prove the effect is to
achieve $\Gamma \sim 0.1\Gamma _n$ at least. As a consequence, it is
possible to define the sensitivity of the experiment. Using $E_c=\hbar
\omega _c$ and $f_r=1/T$, one can write Eq. (\ref{17}) as:
\begin{equation}
\Gamma =K\frac{f_r\tau ^2NI}{E_c^2}g^2  \label{18}
\end{equation}
where $I$ is the intensity of the pulse. In the above expression, $E_c$ is
given in eV, $f_r$ in GHz, $I$ in PW$\cdot $cm$^{-2}$, $\tau $ in fs, and $%
K=g_s^2\mu ^2/(200c\varepsilon _0e^2)\sim 2.74\cdot 10^{-14}$ (in
the relevant units). Figure 4 plots the value of the coupling
constant leading to an observable effect. The values of $g$ are
given against the effective pulsed beam intensity $NI$. Each line is
the lower reachable limit
(such that $\Gamma =0.1\Gamma _n$) for some specific conditions (see Fig. 4).%
\\

(3) \textbf{Conditions of resonance}.

Let us now discuss how to deal with the unknown value of $\eta $ and
its time dependence. The frequency comb source yields simultaneously
a set of frequencies covering a large frequency domain around the
frequency of the carrier wave (Fig. 3). Usual frequency combs can
offer a number of useful frequencies such that $Q\sim 10^5-10^6$
\cite{15,16,17}. As a consequence, for a large frequency domain, one
can simultaneously probe up to a million of frequencies since
neutrons feel them all together. As detailed hereafter, this is a
very efficient way to achieve the resonance without knowledge of the
value of $\eta $, provided it is located in the comb width (see Fig.
3). Of course, as previously suggested, $\eta $ can reach many
possible values in different domains of the electromagnetic
spectrum. As a consequence, in order to increase the chance of
success of such an experiment it could be highly beneficial to
consider different frequency comb sources such as terahertz lasers
\cite{16}, vacuum and extreme ultraviolet lasers \cite{17}, or
free-electron lasers in X-rays domain \cite{25}. With some distinct
frequency combs, one can then expect to quickly cover the most
relevant frequency domains of the electromagnetic spectrum.

As an illustration, let us consider the case $E_c=1$ eV, $\tau =10$ fs, $%
f_r=1$ GHz and $g=10^3$ m$^{-1}$ (see Fig. 4). We suppose that $I=5\cdot
10^{16}$ W$\cdot $cm$^{-2}$ and $N=300$ (see cross in Fig. 4). This leads to
a comb with $Q\sim 6.4\cdot 10^4$ frequencies. The resonance width is $%
\delta \omega \sim 18.6$ Mrad$\cdot $s$^{-1}$. With the values mentioned, we
get $\Gamma \sim 4.11\cdot 10^{-2}$ s$^{-1}$. With such a value, after a
delay of about $2.6$ s, a population of neutrons is reduced from $10$ \%.
Two cases could then be considered:\\

(i) $\eta $ is not time-dependent.

If required, each frequency between two teeth of the comb can be scanned by
varying $\omega _0$ shifting then the frequency $\omega _k$ of each tooth.
Since the resonance width is given by $\delta \omega $, the number of
frequencies to be scanned between two teeth is given by the fineness $F=2\pi
f_r/\delta \omega \sim N$. As a consequence, on the whole frequency domain
covered by the comb, one can probe $NQ$ frequencies just with $N$
experimental runs (usually $N$ can vary from $10$ to $10000$). In the
present illustration, $NQ\sim 1.9\cdot 10^7$ frequencies can be explored.
Spending $2.6$ s on a given set of $Q$ frequencies, for $N=300$, the whole
experiment takes $t_s=13$ min to explore the $NQ$ frequencies. Since we
ignore the right domain of frequencies where $\eta $ locates, in a first
approach, it is not necessary to measure the number of neutrons in the
vessel for each set of $Q$ frequencies.\\

(ii) $\eta $ is time-dependent.

Considering the time variation of $\eta $, we have explained that
the time rate of shift for $\eta $ is likely very small (see section
\ref{freezing}). It means that $\eta $ can spontaneously ''scan''
the frequency comb without a change of $\omega _0$. This would allow
circumventing the problem of time dependence of $\eta $ since a
resonant frequency will be always available provided that $\eta $ is
located into the comb width: a resonance occurs each time that $\eta
$ matches a frequency of a tooth of the comb.
Using the time rate of shift of $\eta $ of $1.7$ meV/day, i.e. about 30 Mrad$%
\cdot $s$^{-1}$/s (see section \ref{freezing}), each resonance is kept
during $0.62$ s since $\delta \omega \sim 18.6$ Mrad$\cdot $s$^{-1}$. For an
effective resonance maintained during $2.6$ s, the resonance frequency must
scan $5$ teeth of the comb at least. Because of the frequency of repetition $f_r$%
, it needs $210$ s to reach the next teeth of the comb, and thus $t_s=14$
min are necessary to scan the five teeth.\\

In both cases, after an experimental run, about $10$ \% of the
neutrons could be transferred to the other brane while about $60$ \%
of the neutrons have disappeared due to the natural neutron decay.
The quantity of neutrons remaining in the vessel can then be counted
while the number of neutron decays is measured during the
experimental run. It is worth noticing that during the whole
experiment, provided that $\eta $ is located on the comb width, the
resonance condition is necessarily encountered one time (if $\eta $
is not time-dependent, when the correct frequency tooth is crossed)
or many times (if $\eta $ is time-dependent, for each crossed
tooth).

One notes that depending on the parameters ($g,$ $\omega _c$, $I$,
$N$, ...)
the required delay of the experiment can be characterized by longer time ($%
t_s\geq 15$ min) or shorter time of storage ($t_s\leq 15$ min) than
in the previous example. In these cases, it could be more relevant
to measure the decay of neutrons (long time of storage) instead of
counting them (short time of storage), although both methods could
be used together. At last, of course, in order to improve the
accuracy of the results, it would be necessary to repeat the same
experiment many times. For $t_s\sim 15$ min, even if the experiment
is repeated $100$ times for instance, a full experimental work takes
about only $1$ day. This makes supposed that our present
experimental framework would lead to fair durations. In return,
though our method allows one to easily achieve the resonance, one
notes that it does not allow one to find the value $\eta $ but
instead a domain of values that corresponds to the width of the
frequency comb (though a dichotomic search using combs with
different widths could allow us to specify the value $\eta$).

\section{Conclusion}

Recent theoretical results have suggested the possibility of matter
swapping between branes \cite{7,8,9}. In the present work, a new
experimental framework has been proposed to demonstrate this
mechanism at a laboratory scale. The experiment, involving a
resonant mechanism, uses an ultracold and polarized neutron gas
interacting with a fine-tuned electromagnetic pulsed radiation. This
experiment relies on the H\"{a}nsch frequency comb technique,
commonly used in spectroscopy to overcome the limitations of the
large bandwidth of electromagnetic sources \cite{14,15,16,17}. The
recorded experimental data are simply the rates at which particle
exchange occurs between branes. Any success in this experiment would
imply major consequences extending far beyond the demonstration of
the existence of other branes. Moreover, the relative simplicity of
the suggested setup is of considerable interest: existing devices
used in spectroscopy \cite {14,15,16,17,25} and neutron physics
investigations (neutron electric dipole moment \cite{10}, Lorentz
invariance \cite{11}, neutron-mirror neutron oscillations \cite{12}
and coupling to axionlike particles \cite{13} for instance) could
possibly match the requirements of a successful experiment. Thus the
present work takes place in a relevant current trend in experimental
works for new physics searches at a low energy scale \cite
{6,10,11,12,13,26}.


\begin{thebibliography}{99}
\bibitem{18}  The gauge field is localized on each domain wall through a
Dvali-Gabadadze-Porrati-Shifman mechanism \cite{2}.
\bibitem{23}  One notes that a magnetic field which would be described by a
function similar to Eq. (\ref{10}) does not give a null
contribution. In addition, it could be shown that such a magnetic
field would not correspond to a relevant magnetic vector potential.
By contrast, the magnetic vector potential here used (Eq.
(\ref{10})) gives a magnetic field (Eq. (\ref {11ante})) that allows
a compensation mechanism that avoids a spin depolarization between
each pulse whatever the laser intensity. As a consequence, the pulse
design must be considered carefully in the experiment \cite{22}.
\end{thebibliography}
\end{document}